\documentclass[%
preprintnumbers,
 amsmath,amssymb,
 aps,
]{revtex4-2}

\usepackage{hyperref}
\usepackage{xcolor}
 \hypersetup{
 colorlinks = true,
 allbordercolors = {white},
 allcolors = {blue},
 }
 
\usepackage{graphicx}
\usepackage{dcolumn}
\usepackage{bm}

\begin{document}

\preprint{APS/123-QED}

\title{Acoustic Equation in a Lossy Medium}

\author{Tapan K. Sengupta}
\affiliation{%
{Department of Mechanical Engineering, IIT (ISM) Dhanbad,}
{Dhanbad,}
{Jharkhand - 826004,}
{India}}
\email{tksengupta@iitism.ac.in}

\author{Prasannabalaji Sundaram}
\affiliation{%
{Department of Aerospace Engineering, IIT Kanpur,}
{Kanpur,}
{Uttar Pradesh - 208016,}
{India}}
\email{prasannabalajis3@gmail.com}

\author{Aditi Sengupta}
\affiliation{%
{Department of Mechanical Engineering, IIT (ISM) Dhanbad, }
{Dhanbad,}
{Jharkhand - 826004,}
{India}}
\email{aditi@iitism.ac.in}

\date{\today}

\begin{abstract}
Here, the acoustic equation for a lossy medium is derived from the first principle from the linearized compressible Navier-Stokes equation without Stokes' hypothesis.  The dispersion relation of the governing equation is obtained, which exhibits both the dispersive and dissipative nature of the acoustic perturbations traveling in a lossy medium, depending upon the length scale. We specifically provide a theoretical cut-off wave number above which the acoustic equation represents a diffusive nature. Such behavior has not been reported before, as per the knowledge of the authors.  

\end{abstract}

\maketitle

\textbf{keywords:} acoustic equation; perturbation pressure; lossy medium; cut-off wavenumber\\
\\
Acoustics, as a branch of science, intrinsically deals with the propagation of signal (information) observed at one point to another closely related signal to another space-time location. Despite a long history of research on wave propagation in this context, there is no clear definition of waves \cite{Whitham74}.  The canonically wave equation, 
\begin{equation} \label{Eq:utt}
    u_{tt} = c^2 u_{xx}
\end{equation}
was first described by D'Alembert \cite{DAlembert1750} in the context of the one-dimensional transverse vibration of string in tension. The solution of Eq.~\eqref{Eq:utt}, subject to initial conditions, can be found in textbooks (see, e.g., \cite{Sengupta13}). This non-dissipative and non-dispersive (i.e., frequency/wavenumber dependency) solution also sets a standard benchmark for developing and calibrating numerical methods in different branches of engineering and applied physics. 

Maxwell \cite{maxwell54, maxwell1865} obtained the wave equations for the electric field E and the magnetic field B, with c as the speed of light (phase speed) in a medium of permeability $\mu_p$ and permittivity $\epsilon_p$ by $c = 1/\sqrt{\mu_p \epsilon_p}$. An electromagnetic wave is transverse in nature, with E and B being perpendicular to wave propagation's direction. 

Some of the other physical phenomena governed by the partial differential equation (PDE) \eqref{Eq:utt} are listed in Mulloth et al. \cite{Mullothetal2015}. Among these are its use in acoustics, Feynman \cite{Feynman65, Feynman69}, elastic wave propagation in solid mechanics \cite{Whitham74} relating applied strain and stress, with the longitudinal displacement $u$, given in Eq.~\eqref{Eq:utt} with $c^2 = E_0/\rho$, where $E_0$ is Young's modulus, and $\rho$ is the density of the medium. 

The present interest in information propagation as sound arises from a desire to develop the acoustic and fluid mechanics description from the first principle for a unified description of disturbance propagation in a lossy medium. In the process, a novel result would be developed for scale-wise propagation of disturbances, following different physical mechanisms without any restriction. 

For sound propagation, the compressible Navier-Stokes equation is mandatory, with the disturbance treated as a small perturbation following the conservation of mass and momentum. In a quiescent, homogeneous medium, one can consider the equilibrium state with the following ansatz: disturbance velocity $\vec V '$ and disturbance density $\rho '$ develops with no mean motion ($\vec V = 0$) and a steady state for the unperturbed density $\left( \frac{\partial \bar \rho}{\partial t} = 0\right)$. 

Conservation of mass: 
\begin{equation} \label{Eq:CE}
    \frac{\partial \rho}{\partial t} + \nabla \cdot \rho \vec{V} = 0 
\end{equation}
Conservation of momentum equation without body forces: 
\begin{equation} \label{Eq:ME}
\rho \left( \frac{\partial \vec V}{\partial t} + \left( \vec V \cdot \nabla\right) \vec V \right)  = -\nabla p + \nabla \cdot \left( \lambda \left(\nabla \cdot \vec V \right) \bf{I} \right) + \nabla \cdot \left[ \mu \left( \nabla \vec V + \nabla \vec V ^T \right)\right]
\end{equation}

Here, $\bf{I}$ is an identity matrix with rank three. Consider acoustic signal as a small perturbation over the mean flow so that the velocity, the density, and the pressure are expressed as 

\begin{equation}
\vec V = \bar V      + \epsilon \vec V ' ; ~~~\rho     = \bar \rho + \epsilon \rho ' ;~~~ p          = \bar p       + \epsilon        p '  \label{Eq:separation}
\end{equation}
Next, consider the propagation of the acoustic signal in a quiescent flow (i.e., $\vec V = \vec 0$) in a homogeneous medium (i.e, constant $\bar \rho$), so that, 
\begin{equation} \label{Eq:04}
\vec V = \epsilon \vec V'.
\end{equation}
The $O(\epsilon)$ equation resulting from Eq.~\eqref{Eq:CE} yields, 
\begin{equation} \label{Eq:05}
    \frac{\partial \rho '}{\partial t} + \bar \rho  \nabla \cdot \vec{V}' = 0 
\end{equation}	The $O(\epsilon)$ equation resulting from Eq.~\eqref{Eq:ME} yields, 
\begin{equation} \label{Eq:06}
\bar \rho  \frac{\partial \vec V '}{\partial t}  = -\nabla p' + \nabla \cdot  \left( \lambda \left(\nabla \cdot \vec V'  \right)\bf{I} \right) + \nabla \cdot \left[ \mu \left( \nabla \vec V' + \nabla \vec V^{' T} \right)\right]
\end{equation}
From Eq.~\eqref{Eq:05}: $\nabla \cdot \vec V ' = \frac{-1}{\bar \rho} \frac{\partial \rho '}{\partial t}$ and differentiating with respect to time yields, 
\begin{equation} \label{Eq:07}
\frac{\partial}{\partial t} \left(  \nabla \cdot \vec V '\right) = \frac{-1}{\bar \rho} \frac{\partial^2 \rho'}{\partial t ^2}
\end{equation}
Taking divergence of Eq.~\eqref{Eq:06}, one gets
\begin{equation} \label{Eq:08}
\bar \rho \frac{\partial}{\partial t} \left(  \nabla \cdot \vec V '\right)  = -\nabla^2 p' + \lambda \nabla^2 \left( \nabla \cdot \vec V' \right) +  2\mu  \nabla^2 \left(\nabla \cdot \vec V'\right)
\end{equation}
From Eq.~\eqref{Eq:07} one gets, 
\begin{equation} \label{Eq:09}
-\frac{\partial^2 \rho'}{\partial t^2}  = -\nabla^2 p' - \left( \lambda + 2 \mu\right) \nabla^2 \left( \frac{1}{\bar \rho} \frac{\partial \rho'}{\partial t} \right) 
\end{equation}
From the polytropic relation, we get, 
\begin{equation} \label{Eq:10}
	\frac{\partial \rho'}{\partial t} = \frac{1}{c^2}\frac{\partial p'}{\partial t}
\end{equation}
Therefore, 
\begin{equation} \label{Eq:11}
\frac{1}{c^2}\frac{\partial^2 p'}{\partial t^2}  =\nabla^2 p' + \frac{\lambda + 2 \mu}{\bar \rho c^2}  \nabla^2 \left( \frac{\partial p'}{\partial t} \right) 
\end{equation}
\begin{equation} \label{Eq:12}
\frac{\partial^2 p'}{\partial t^2}  = c^2\nabla^2 p' + \nu \frac{\partial }{\partial t} \nabla^2  p'
\end{equation}
Here, $\nu = \frac{\lambda + 2 \mu}{\bar \rho}$, i.e., Stokes' hypothesis \cite{stokes1851} is not invoked.

\subsection{Characteristics of Acoustic Equation}
The acoustic equation derived in the lossy medium is distinctively different from the one derived by Feynman \cite{Feynman65, Feynman69} for loss-less medium. Equation~\eqref{Eq:12} without loss term ($\nu =0$) is the classical wave equation as described before \cite{Whitham74,DAlembert1750,Sengupta13,maxwell1865,maxwell54,Mullothetal2015}. In contrast to the wave equation (being a hyperbolic PDE), the present acoustic equation needs characterization. As the governing conservation equation is similar to the Navier-Stokes equation, this is adequately investigated by the global spectral analysis \cite{Sengupta13, SumanSengPrasMoha17}. 

To elucidate the fundamentals, attention is focused here on Eq.~\eqref{Eq:12} for the one-dimensional governing equation \cite{Blackstock2000} by, 
\begin{equation} \label{Eq:13}
\frac{\partial^2 p'}{\partial t^2}  - c^2\frac{\partial^2 p'}{\partial x^2} - \nu \frac{\partial^3  p' }{\partial t\partial x^2}  = 0
\end{equation}

The hydrodynamic and acoustic events occur on disparate scales, and it remains as a challenge to simultaneously solve flow and acoustic problems. For the purpose of analysis, represent the fluctuating pressure by, 
\begin{equation} \label{Eq:14}
p'(x,t) = \int \int \hat p (k,\omega)e^{\iota (kx-\omega t)} dkdw
\end{equation}
Rewriting Eq.\eqref{Eq:13} in the spectral plane by using the above representation one gets the quadratic dispersion relation as, 
\begin{equation} \label{Eq:15}
\omega ^2  + \iota \nu k^2 \omega - c^2 k^2 = 0 
\end{equation}
This yields the solution as 
\begin{equation}\label{Eq:16}
\omega_{1,2} = \frac{-\iota \nu k^2}{2} \pm kcf
\end{equation}
Where, $f = \sqrt{1-\left(\frac{\nu k}{2c} \right)^2}$. Treating the wavenumber $k$, as the independent variable, the dispersion relation in Eq.~\eqref{Eq:16}, provides the following amplification factor as, 
\begin{equation}
G_{1,2} = e^{-\iota \omega_{1,2} dt}
\end{equation}
for an incremental timestep $dt$ for $f > 0$. These complex quantities indicate phase shifts given by, 
\begin{equation}\label{Eq:19}
\beta_{1,2} = \pm kcf~dt
\end{equation}
Thus the positive value of $f$ indicates the action of a lossy medium in contrast to the non-dispersive nature of the classical  wave equation. Apart from the dispersive nature of the acoustic equation, there is another aspect of this equation in propagating the fluctuating pressure, which is described next.  

\subsection{Wavenumber dependence of acoustic equation}

One of the central results of the acoustic equation is the non-dispersive nature of the lossy medium for the propagation of fluctuating pressure. From Eq.\eqref{Eq:16}, this is evident. One also notes the ultraviolet range of $k \rightarrow \infty$, when one can approximate, $f = \iota |f|$ and then $\omega_{1,2}$ both becomes pure imaginary and that can explain the absence of ultraviolet catastrophes that have been explained for electromagnetic radiation \cite{ehrenfest1911zuge}. 

However, there is another possibility of a qualitative change in the characteristics of the acoustic equation itself. One notices a cut-off wavenumber ($k_c = 2\bar \rho c/(\lambda + 2\mu)$) for which $f\equiv 0$, above which $\omega_{1,2}$ becoming strictly imaginary, implying the acoustic equation to represent a diffusion equation. This is a novel result that shows the small perturbation in fluid flow and in acoustics to be given by the fluctuating pressure which displays wavy nature for wavenumbers lower than $k_c$, and the wavenumbers above $k_c$ become diffusive.

\end{document}